\documentstyle[aps,floats,prl]{revtex}
\tighten

\begin{document}
\draft

\twocolumn[\hsize\textwidth\columnwidth\hsize\csname @twocolumnfalse\endcsname

\title{Initial data for two Kerr-like black holes}

\author{Sergio Dain}

\address{Albert-Einstein-Institut,
Max-Planck-Institut f{\"u}r Gravitationsphysik,
Am M\"uhlenberg 1, D-14476 Golm, Germany}

\date{\today}

\maketitle
\begin{abstract}
We prove  the existence of a family of initial data for the Einstein
vacuum equation which can be interpreted as the data for two Kerr-like
black holes in arbitrary location and with spin in arbitrary
direction. 
When the mass parameter of one of them
is zero, this family reduces exactly to the Kerr initial data.  The
existence proof is based on a general property of the Kerr metric which
can be used in other constructions as well. Further generalizations are
also discussed. 
\end{abstract}


\vskip0.5pc]

\emph{Introduction.}
---Black-hole collisions  are considered as one of the most important
sources of gravitational radiation that may be observable with the
gravitational wave detectors currently under construction. The first
step in the study of a black-hole collisions is to provide proper initial
data for the Einstein vacuum equation.  
Initial data for two black holes were
first constructed by Misner\cite{Misner60}, shortly  after Brill and
Lindquist\cite{Brill63} studied  a similar data but with a different
topology which considerable simplifies  the construction.
Bowen-York\cite{York} included  linear and angular  momentum. 
 Generalizations of these data were studied  in
\cite{Beig}\cite{Beig94} \cite{Beig00} \cite{Brandt97b} (see also the review
\cite{Cook00} and the references therein). 

These families of initial data depend on the mass, the momentum, the spin,
and the location of each black hole. When the mass parameter of one of
the holes is zero, one obtains initial data for only one black hole. It
is physically reasonable to require that these black hole data be
stationary, i.e.; a slice of Schwarzschild or Kerr space time. If this
is not the case, it means that spurious gravitational radiation is
present in the initial data.  In the case of \cite{Misner60} and
\cite{Brill63}, this  gives the Schwarzschild initial data.  But for the
other cases, where angular momentum is included, one does not obtain the
initial data of the Kerr metric.

The Kerr initial data are not included in the families considered above 
because the restrictions imposed on the conformal 3-metric are too
strong. In most cases it is assumed that the conformal metric is flat.
However, it appears that the Kerr metric admits no conformally flat
slices (in fact, in \cite{Price00} it has been shown that there does not
exist  axisymmetric, conformally flat foliations  of the Kerr
spacetime  that smoothly reduce, in the Schwarzschild limit, to slices of 
constant Schwarzschild time). Much weaker
conditions have been imposed on the 3-metric which nevertheless exclude
Kerr data; for example in \cite{Beig} it is required that the conformal
metric admit a smooth compactification. We will see that, at least for
the Boyer-Lindquist slices, this condition is also strong enough to
exclude Kerr data.  

The purpose of this article is to generalize these constructions above in
order to include the Kerr initial data as a particular case in which the
mass parameter of one black hole is zero.  This family of initial data is
the natural generalization of the one found in \cite{Brill63}. 
For the axisymmetric case, similar data have been calculated numerically
\cite{Price98}. Recently, a  different type of initial data has been
also calculated 
numerically  \cite{Marronetti00}. 
The existence proof is based on general property  of
the Kerr initial data, which may also be useful in other constructions.   
The plan of this article is as follows: first we prove this key property
of the Kerr metric;  second we give a remarkably simple application of
it in the construction of data that can be interpreted as data for a 
Schwarzschild and a Kerr black hole; then we construct more complicate
initial data, which can be interpreted as the data for two Kerr black
holes. Finally, certain generalizations are discussed.

\emph{The Kerr initial data.}
---Consider the Kerr metric in the Boyer-Lindquist
coordinates $(t, \tilde r, \vartheta, \phi )$\cite{Kerr63}
\cite{Boyer67}, with mass
$m$ and angular momentum $a$ such that $m>a$. 
Take any  slice $t=const$. Denote by $\tilde h^k_{ab}$ the intrinsic three
metric of the slice and by $\tilde \Psi_k^{ab}$ its  extrinsic curvature. 
These  slices are maximal, i. e. 
$\tilde h^k_{ab}\,\tilde \Psi_k^{ab} = 0$. 
The metric $\tilde h^k_{ab}$ is given in the
coordinates $(\tilde r, \vartheta, \phi )$ by
\begin{equation}
\label{eq:ptildehBL}
\tilde h^k \equiv \frac{\Sigma}{\Delta} d\tilde r^2
+ \Sigma d\vartheta^2 +\eta  d \phi^2, 
\end{equation}
where
\begin{equation}
\label{eq:Sigma}
\Sigma=\tilde r^2+a^2 \cos^2 \vartheta, \quad \Delta =\tilde r^2+a^2-2m\tilde r,
\end{equation}
and
\begin{equation}
  \label{eq:eta}
  \eta=\sin^2\theta (\Sigma +a^2 \sin^2\theta (1+\hat\sigma)), \quad \hat
  \sigma=\frac{2m\tilde r}{\Sigma}.
\end{equation}
The metric is singular where $\Delta$ or $\Sigma$ vanishes. The zeros
of the function $\Delta$ are given by 
\begin{equation}
\label{eq:r+-}
\tilde r_+=m+\delta, \quad \tilde r_-=m-\delta,  
\end{equation}
with $\delta=\sqrt{m^2-a^2}$.

Consider the coordinate transformation
\begin{equation}
\label{eq:itrpsi}
 \tilde r =  \frac{\alpha^2 \cos^2(\psi/2) +\delta^2
 \sin^2(\psi/2)}{\alpha\sin \psi} +m, \quad 0\leq \psi \leq  \pi,
\end{equation}
where $\alpha$ is a positive constant.
This transformation is the composition of the transformation to the quasi
isotropical radius $\bar r$ and a stereographic projection, i.e.
\begin{equation}
\label{eq:ipsi}
\tilde r = \bar r +m+\frac{\delta^2}{4\bar r}, \quad 
\bar r = \frac{\alpha\cos (\psi/2)}{2\sin (\psi/2)}.
\end{equation}
It  is defined for  for $\tilde r>\tilde
r_+$, and  becomes singular at $\tilde r_+$. 
We consider $(\psi, \vartheta, \phi)$ as standard coordinates on 
$S^3$.  The south pole is given by $\psi=0$ and the north pole by
$\psi=\pi$, we will denote them by $\{0\}$ and $\{\pi\}$ respectively.  
Due to the isometry 
$\bar r \rightarrow \delta^2/(4\bar r)$, the transformation 
(\ref{eq:itrpsi}) maps one copy of the region $\tilde r>\tilde
r_+$ into the region $\psi >\psi_+$ of $S^3$, where  
$\psi_+=2\arctan(\alpha/\delta)$,  and
another copy into $\psi <\psi_+$.  
In the new coordinates the metric (\ref{eq:ptildehBL}) 
extend to a smooth metric in $S^3-\{0\}-\{\pi\}$. This manifold
defines a space like hypersurface in the Kerr space time which, in 
figure 28 of \cite{Hawking73}, can be indicated by horizontal straight line
going from one apex of a region I to the opposite apex of the adjacent
region I. The poles $\{0\}$ and $\{\pi\}$ are precisely these apexes, they
represent the space like infinities of the initial data. 
This hypersurface is a Cauchy surface for an asymptotically flat region
of the Kerr space time (comprising two regions I and II respectively). 

Using the conformal factor
\begin{equation}
  \label{eq:iTheta2}
  \theta_k =\frac{\Sigma ^{1/4}}{\sqrt{\sin \psi}},
\end{equation}
define the conformal metric $h^k_{ab}$ by
\begin{equation}
  \label{eq:ihth}
  h^k_{ab}=\theta_k ^{-4} \tilde h^k_{ab}. 
\end{equation}
The conformal factor $\theta_k$ is singular at $\{0\}$ and $\{\pi\}$,
\begin{equation}
\label{eq:bThetaK}
\lim_{\psi\rightarrow \pi}(\psi-\pi)\theta_k
=\sqrt{\delta}, \quad \lim_{\psi\rightarrow 0}\psi \theta_k=\sqrt{\alpha}.  
\end{equation}
The
metric $h^k_{ab}$ has the form
\begin{equation}
  \label{eq:hk}
h^k_{ab}=h^0_{ab}+a^2  f v_av_b,  
\end{equation}
where $h^0_{ab}$ is the standard metric of  $S^3$, the smooth vector
field 
$v_a$ is given by $v_a\equiv \sin^2\psi \sin^2\vartheta (d\phi)_a$, and the
function $f$, which contains the non-trivial part of the metric, is
given by 
\begin{equation}
  \label{eq:phatf}
  f =\frac{(1+\hat \sigma)}{\Sigma\,\sin^2 \psi}.
\end{equation}
The function $f$ depends on $a,m, \sin\psi, \cos\vartheta$. It 
is smooth in $S^3-\{0\}-\{\pi\}$. In order to analyze the differentiability of 
$f$ at the poles, take a normal coordinate system $x^i$ with respect
to the metric $h^k_{ab}$, centered at one of
the poles, define the radius $|x| = (\sum_{i=1}^3 (x^j)^2 )^{1/2}$. 
In terms of these coordinates the function $\psi$, given
by $\psi=|x|$, is seen to be a $C^\alpha$ function of $x^i$. From the
expression (\ref{eq:phatf}) one can prove that the function $f$  has the 
form 
\begin{equation}
\label{eq:ihatf}
 f =f_1 +f_2 \sin^3 \psi  ,
\end{equation}
where $ f_1$ and $ f_2$ are smooth functions in the
neighborhood of the poles, with respect to  the  coordinates $x^i$. 
Since $\sin^3 \psi \in W^{4,p}$, $p<3$, (see e.g. \cite{Adams} for the
definitions of the Sobolev and H\"older spaces $W^{s,p}$ and
$C^{m,\alpha}$) from expression (\ref{eq:ihatf})
we see  that
\begin{equation}
  \label{eq:sob}
  h^k_{ab} \in W^{4,p}(S^3), \quad p<3.
\end{equation}

This is the crucial property of the metric that will be used in the
existence proof. In fact, it is the only property of the Kerr metric
that we will need. It implies, in particular, that the
metric is in $C^{2,\alpha}(S^3)$. Since the poles $\{0\}$ and
$\{\pi\}$ are the infinities of the data, the expression (\ref{eq:ihatf})
characterizes the fall-off behavior of the Kerr initial data near space
like infinity. The Ricci scalar $R$ of the metric
$h^k_{ab}$ is a continuous function of the parameter $a$, and for
$a=0$ we have that $R=6$, the scalar curvature of $h^0_{ab}$.
Thus, if $a$ is sufficiently small, $R$ will be a positive function on
$S^3$. In the following we will assume the latter condition to be
satisfied. 

It remains to analyze the extrinsic curvature of the Kerr initial data. 
Define  $\Psi_k^{ab}$ by  
\begin{equation}
 \label{eq:PsiKerrc}
\Psi_k^{ab}=\theta_k^{10} \tilde \Psi_k^{ab}.
\end{equation}
The tensor $\Psi_k^{ab}$ is smooth in  $S^3-\{0\}-\{\pi\}$ and at the
poles it has the form
\begin{equation}
  \label{eq:PsiKerr}
 \Psi_k^{ab}=\Psi^{ab}_{J}+Q^{ab} 
\end{equation}
where $\Psi^{ab}_{J}=O(|x|^{-3})$ and it is  trace-free and divergence
free with respect to the flat metric (it contains the angular momentum of 
the data and the explicit form of this tensor is given in
\cite{York}).  The tensor $Q^{ab}$ is $O(|x|^{-1})$. If $a=0$ then
$\Psi_k^{ab}=0$.

The coordinate transformation (\ref{eq:itrpsi}) simplifies
considerably if we choose $\alpha=\delta$. This choice makes the
metric (\ref{eq:hk}) symmetric with respect to $\psi=\pi/2$, it is
useful in explicit calculations. Nevertheless, this choice is
inconvenient for our present purpose, since it is singular when
$\delta=0$  and we want to have the flat initial data in this limit. In the
following we will assume $\alpha=1$.

\emph{Initial data with Schwarzschild-like and Kerr-like asymptotic
ends} ---The  conformal approach to find solutions of the constraint
equations with many asymptotically flat end points $i_n$ is the
following (cf. \cite{Choquet99}, \cite{Choquet80} and the reference
given there.  The setting outlined here, where we have to solve
(\ref{Lich}), (\ref{thetai}) on the compact manifold  has been studied in
\cite{Beig}, \cite{Friedrich88},
\cite{Friedrich98}). Let $S$ be a compact manifold (in our case it will be
$S^3$), denote by $i_n$ a finite number of points in $S$, and define the
manifold $\tilde S$ by 
$\tilde S = S \setminus \bigcup i_n$. We assume that
$h_{ab}$ is a positive definite metric on $S$, with covariant derivative
$D_a$, and 
$\Psi^{ab}$ is  a trace-free symmetric tensor, which satisfies   
\begin{equation} 
\label{diver}
D_a \Psi^{ab}=0 \quad\mbox{on}\quad \tilde S.
\end{equation}
Let $\theta$ a solution of 
\begin{equation} 
\label{Lich}
L_h \theta=-\frac{1}{8}\Psi_{ab}\Psi^{ab}\theta^{-7}
\quad \mbox{on}\quad \tilde S , 
\end{equation}
where   $L_h=D^aD_a-R/8$.
Then the physical fields  $(\tilde h, \tilde
\Psi)$ defined by  $\tilde{h}_{ab} =
\theta^4 h_{ab}$ and $\tilde{\Psi}^{ab} = \theta^{-10}\Psi^{ab}$ will
satisfy the vacuum constraint equations on $\tilde S$. To ensure
asymptotic flatness of the data at the points $i_n$ we require at
each point $i_n$
\begin{equation} 
\label{Psii}
\Psi^{ab} = O(|x|^{-4}) \quad\mbox{as}\quad x \rightarrow 0,  
\end{equation}
\begin{equation} 
\label{thetai}
\lim_{|x|\rightarrow 0} |x|\theta = c_n,
\end{equation}
where the $c_n$ are positive constants, and $x^i$ are normal coordinates
centered at $i_n$. 

Since $(h^k,\Psi_k)$ are obtained from the Kerr solution, they satisfy
equations (\ref{diver}) and (\ref{Lich}); and also the boundary
conditions (\ref{Psii}) and (\ref{thetai}) at  each of the poles,
since they satisfy equations (\ref{eq:bThetaK}) and (\ref{eq:PsiKerr}). 
The Kerr metric $h^k$ satisfies (\ref{eq:sob}), then the coefficients
of the elliptic operator $L_{h^k}$ satisfy the hypothesis of 
 the existence theorems proved in \cite{Dain99},
in particular they are in $C^\alpha(S^3)$. 
 From these theorems, it follows that,
for arbitrarily chosen point $i\in S^3$, there exists a unique, positive
function $\theta_i$, which satisfies
\begin{equation}
  \label{eq:Green}
L_{h^k} \theta_i=0, \text{ in } S^3-\{i\},  
\end{equation}
and at $i$
\begin{equation}
  \label{eq:bGreen}
\lim_{x\rightarrow 0} |x|  \theta_i=1, 
\end{equation}
$|x|$ denoting the distance from $i$.

We denote by $\theta_0$, $\theta_\pi$ the solutions so obtained by
choosing the point $i$ to be $\{0\}$ and $\{\pi\}$ respectively
and write the  Kerr conformal factor $\theta_k$ in the form 
\[
\theta_k=\theta_0+ \sqrt{\delta}\theta_\pi+u_k.
\]
The function $u_k$ is then in $C^\alpha(S^3)$. 

In order to produce another asymptotic end in the initial data above,
take an arbitrary point $i_1\in S^3$, different from $\{0\}$ and
$\{\pi\}$, with coordinates $(\psi_1, \vartheta_1, \phi_1)$, and consider
the corresponding function $\theta_1$ which satisfies (\ref{eq:Green})
and (\ref{eq:bGreen}) in $i_1$.  Define the function $\theta_{sk}$ by 
\begin{equation}
  \label{eq:ThetaSK}
\theta_{sk}=\theta_0+\sqrt{\delta}\theta_\pi+
\sqrt{m_1} \theta_1\sin (\psi_1/2)+u_{sk},
\end{equation}
where $m_1$ is an arbitrary, positive, constant.   Insert this in
equation  (\ref{Lich}), where we use $\Psi_k^{ab}$ in place of
$\Psi^{ab}$ and $h^k_{ab}$ in place of $h_{ab}$. Observing that we used
(\ref{eq:Green}), we obtain an equation for $u_{sk}$ on $S^3$. 
 The right hand side of this 
equation is in $L^2(S^3)$, since the singular behavior of $\Psi^{ab}$ 
at the poles is canceled by the negative power  of $\theta$. 
 In \cite{Dain99} it has been proven that this equation has a unique,
positive, solution, under the conditions stated above.  In
this particular case, since there is no linear momentum in any of  the
asymptotic ends, the extrinsic curvature  $\Psi_k^{ab}$ is $O(|x|^{-3})$ at the
poles, and then  the right hand side of (\ref{Lich}) is in
$C^\alpha(S^3)$. But it is important to recall that the existence theorem also applies 
when a  term with  linear momentum is  included at the ends.  We
will come back to this point later on. 
It follows that,
for arbitrarily chosen point $i_1$, the tensors 
\begin{equation}
  \label{eq:Skini}
  \tilde h^{sk}_{ab}=\theta^{-4}_{sk} h^k_{ab}, \quad \tilde \Psi_{sk}^{ab} = 
  \theta^{10}_{sk} \Psi_{k}^{ab},
\end{equation}
define a solution of the vacuum constraint equations. 

These initial data have three asymptotic ends $\{0\}$, $\{\pi\}$ and
$\{i_1\}$, i.e.; they have the same topology as the data obtained in
\cite{Brill63}. Moreover, when  $a=0$ we obtain exactly the same solution
obtained in \cite{Brill63} with mass $m$ and $m_1$ (this is the
reason for the factor $\sin (\psi_1/2)$ in (\ref{eq:ThetaSK})). Then, at least for small $a$, we expect the same
behavior of the apparent horizons as the one discussed there. That
is, when the mass parameter $m$ and $m_1$ are small with respect of
the separations between the  ends, only
two  apparent horizons will appear,  surrounding  $\{\pi\}$ and
$\{i_1\}$. This makes a geometric distinction between the ends
$\{\pi\}$ and $\{i_1\}$, which have an apparent horizon around them, and
$\{0\}$, which has not. When the separation of the ends is comparable
with  the masses, we expect that another apparent horizon appears around
$\{0\}$.  The evolution of these data  will
presumably contain an event horizon, the final picture  of the whole
space time will be similar to the one shown in figure 60 of
\cite{Hawking73}, which represents a collision and merging of two black
holes.  The asymptotic end $\{i_1\}$  can be interpreted as the
`Schwarzschild' end, since when $m=a=0$ (i.e.; when the end $\{\pi\}$ 
is not present) we obtain exactly the Schwarzschild initial data with mass
$m_1$. One can expect that the geometry near $\{i_1\}$ approximates, in
some sense, the Schwarzschild geometry. 
In an analogous way, when $m_1=0$, we obtain the Kerr initial data with
mass $m$ and angular momentum $a$. We then say that $\{\pi\}$ is the
`Kerr' asymptotic end. If we chose $\vartheta_1 \neq
0,\pi$ the data will be non-axially symmetric. It is remarkable that, in
order to construct these data, the only new function that one has to
compute is the conformal factor $\theta_{sk}$, the conformal metric
(\ref{eq:ihth}) and the conformal extrinsic curvature (\ref{eq:PsiKerrc})
being given explicity by the Kerr geometry. 

\emph{Initial data with two Kerr-like asymptotic ends.}
---Take the Kerr initial data in coordinates $(\bar r, \vartheta,
\phi)$.  Make a rigid rotation such that the spin  point in the
direction of an arbitrary unit vector $S_1^a$, and make
  a shift of the origin $\bar r=0$ to the  coordinate  position of an
arbitrary point $i_1$. Let the mass and the modulus of angular momentum of this
data  be $m_1$ and $a_1$.  
We  apply the stereographic projection 
 (\ref{eq:ipsi}) and the conformal rescaling
(\ref{eq:iTheta2}).   Then, we  obtain a rescaled metric
$h^{k_1}_{ab}=h^0_{ab}+ a^2_1 f_1 v^1_av^1_b$, where  $f_1$
and $v^1_a$ are obtained from $f$ and $v_a$ by the rotation and the shift of the origin,
 they depend on the coordinates of the point $i_1$ and the vector
 $S_1^a$.  In $S^3$, this coordinate transformation is a smooth conformal
 mapping with a fixed  point at $\{0\}$.  
In an analogous
way we define the corresponding rescaled extrinsic curvature
$\Psi_{k_1}^{ab}$. Take another vector $S_2^a$ and another 
point $i_2$ and make the same construction.   We define the
following metric 
\begin{equation}
  \label{eq:hp}
h^{kk}_{ab}=h^0_{ab}+ a^2_1  f_1 v^1_av^1_b +a^2_2  f_2 v^2_av^2_b. 
\end{equation}
By (\ref{eq:sob}) we have that this metric is in $W^{4,p}(S^3)$.  Is also clear that for small $a_1$ and $a_2$ the scalar 
curvature is positive. 

Set  $\bar \Psi_{k_1}^{ab}$ to be the trace-free part of
$\Psi_{k_1}^{ab}$ with respect to the metric $h^{kk}_{ab}$. Define the 
tensor $\Psi^{ab}_{kk}$ by 
\begin{equation}
  \label{eq:PsiKK}
  \Psi^{ab}_{kk}=\bar \Psi^{ab}_{k_1}+\bar \Psi^{ab}_{k_2}+(l w)^{ab},
\end{equation}
where $(l w)^{ab}$ is the conformal Killing operator $l$, with
respect to the metric $h^{kk}_{ab}$, acting
on a vector $w^a$.  In
\cite{Dain99} we have proved that there exist a unique $w^a\in W^{2,p}(S^3)$ such
$\Psi_{ab}^{kk}$ satisfies  (\ref{diver}).  When $a_1$ or $a_2$ is
equal to zero, then $\Psi^{ab}_{kk}$ is equal to  $\Psi_{k_1}^{ab}$ or 
$\Psi_{k_2}^{ab}$ respectively, since the solution is unique.

Define the conformal factor $\theta_{kk}$ by 
\begin{equation}
  \label{eq:ThetaKK}
  \theta_{kk}=\theta_0+\sqrt{\delta_1}\theta_1\sin (\psi_1/2)+\sqrt{\delta_2}\theta_2 \sin (\psi_2/2) +u_{kk},
\end{equation}
where $\theta_1$ and $\theta_2$ satisfy (\ref{eq:Green}) and
(\ref{eq:bGreen}) for  $i_1$ and $i_2$ respectively, with respect to
the metric (\ref{eq:hp}). Using again the
existence theorem proved in \cite{Dain99}, we have that 
there  exist a unique, positive,  solution $u_{kk}$ of  equation
(\ref{Lich}), where we have replaced $h_{ab}$ by $h^{kk}_{ab}$ and
$\Psi^{ab}$ by $\Psi_{kk}^{ab}$.
Then, we have constructed a solution 
\begin{equation}
  \label{eq:KKini}
  \tilde h^{kk}_{ab}=\theta^{-4}_{kk} h^{kk}_{ab}, \quad \tilde \Psi_{kk}^{ab} = \theta^{10}_{kk} \Psi^{ab}_{kk}.
\end{equation}
of the constraint equation. This solution   has also three asymptotic ends $\{0\}$,
$\{i_1\}$ and $\{i_2\}$, and when $a_1=a_2=0$ we obtain the solution
\cite{Brill63} with mass $m_1$ and $m_2$.  Then, at least for small
$a_1$ and $a_2$,  we expect that the same behavior of the apparent
horizons as the one discussed before. The main difference is that now 
 both ends $i_1$ and $i_2$ are `Kerr'
ends, since when $m_1=a_1=0$ we obtain the Kerr initial data, and the
same is true for $m_2=a_2=0$.  We can expect that the geometry
near each  of this ends is similar, in some sense,  to the geometry of
the Kerr 
initial data, when the mass are small with respect to the
separation. Numerical comparison for the conformal factor, which exhibits
this behavior, has been made  in \cite{Price98} for the
axisymmetric case.  

\emph{Conclusion.}
We have constructed a family of initial data that can be interpreted
as representing   two Kerr black holes. It reduces exactly to the Kerr
initial data when the mass  of one of them is
zero.  This is the first rigorous
proof of the existence of such a class of initial data.
 We have chosen the ansatz
(\ref{eq:hp}), which is perhaps the simplest one, but other choices are
possible too. The only  requirement we must impose on
the conformal  metric (\ref{eq:hp}) is that it reduces to the conformal
Kerr metric when $a_1$ or
$a_2$ is equal to zero and satisfies (\ref{eq:sob}), this is a very
mild condition. It is also possible
 to add an extra term in the extrinsic curvature (\ref{eq:PsiKK})
which contains the linear momentum of each black hole. The existence
proof is exactly the same  (see \cite{Dain99}).  However, we will not
have either Kerr or Schwarzschild when only one black hole is present, 
 since the
Boyer-Lindquist slices  are
not boosted,  this is  exactly the same situation 
as for the boosted data given in \cite{York}. The conformal Kerr
metric has the    special form (\ref{eq:ihatf}), this can be used to prove
additional regularity properties of the initial data, this will be
done in  future work. In order to see whether the gravitational
waves emitted in the case of our data differ in a significant way
from the waves observed for Bowen-York data, it would be interesting to
compare the numerical evolution of the corresponding space-times.

I would like to thank  J. Baker, B. Br\"ugmann, M. Campanelli, S. Husa, C. Lousto, R. Price, J. Pullin for discussions, and especially  H. Friedrich for a carefully reading of the manuscript.

\end{document}